\documentstyle[aps,preprint]{revtex}
\tighten

\begin{document}

\title{ Flavor mixing and time of flight delay 
of supernova neutrinos} 

\author{\it Sandhya Choubey\thanks{e-mail: sandhya@tnp.saha.ernet.in} 
and Kamales Kar\thanks{e-mail: kamales@tnp.saha.ernet.in}\\ 
Saha Institute of Nuclear Physics,\\1/AF, Bidhannagar,
Calcutta 700064, INDIA.}
\maketitle

\def\nue{{\nu_e}}
\def\anue{{\bar\nu_e}}
\def\numu{{\nu_{\mu}}}
\def\anumu{{\bar\nu_{\mu}}}
\def\nutau{{\nu_{\tau}}}
\def\anutau{{\bar\nu_{\tau}}}
\newcommand{\dm}{\mbox{$\Delta{m}^{2}$~}}
\newcommand{\st}{\mbox{$\sin^{2}2\theta$~}}

\begin{abstract}

The neutrinos from galactic supernovae can be detected by the 
Sudbury Neutrino Observatory and the Super-Kamiokande. The effect 
of neutrino mass can show up in the observed neutrino signal by 
(i) delay in the time of flight and (ii) distortion of the neutrino 
energy spectrum due to flavor mixing. We discuss a combination of the 
two effects for both charged and neutral current processes in the 
detectors for realistic three flavor scenarios and show that 
neutrino flavor mixing can lead to non-trivial changes in the event rate 
as a function of time. 

\end{abstract}
 
\newpage
\section{Introduction}

The subject of neutrino astronomy was firmly established in February 
1987 with the detection of neutrinos coming from the explosion of 
SN1987A in the large magellanic cloud. Ever since there have been lot of 
activities in this field. Particularly neutrino mass and its implications 
on supernova neutrino detection have been issues of much discussion. 
Nonzero neutrino mass was conjectured long ago as a plausible solution to 
the solar neutrino deficit problem \cite{solar} and the atmospheric 
neutrino anomaly \cite{atm,skatm}. The result from the Super-Kamiokande 
atmospheric neutrino experiment has finally confirmed that 
neutrinos are indeed massive \cite{skatm}. Among the terrestrial 
accelerator/reactor experiments, only the LSND in Los Alamos has claimed 
to have seen neutrino mass and mixing \cite{lsnd}.

All the three above mentioned evidence for neutrino mass are in a way 
indirect as they come from neutrino mixing. The direct kinematical 
measurements of neutrino mass in laboratory experiments at present 
give extremely poor limits \cite{balasev,lab} which far exceeds the 
cosmological 
bound \cite{raf}. The other more promising possibility of determining 
$\numu/\nutau$ masses is through the observation of neutrinos from 
stellar collapse. 

The core of a massive star $(M\ge 8M_\odot)$ starts collapsing 
once it runs out of nuclear fuel.
The collapse continues to densities beyond the nuclear
matter density after which a bouncing of the infalling matter takes place
leading to supernova explosion and the formation of a protoneutron star.
Only a small fraction of the huge gravitational energy released in the process
goes into the explosion and the rest of the energy is carried away by
neutrinos and antineutrinos of all three flavors. These neutrinos for
galactic supernova events can be detected by detectors like the Sudbury 
Neutrino Observatory (SNO) and the Super-Kamiokande (SK). 
The effect of neutrino mass can show up in the observed neutrino 
signal in these detectors in two ways, 
\begin{itemize}
\item by causing delay in the time of flight measurements
\item by modifying the neutrino spectra through neutrino flavor mixing
\end {itemize}

For neutrinos traveling distances $\approx$ 10 kpc, even a small mass 
will result in a measurable delay in the arrival time 
\cite{bkg,krauss,bv} resulting in the distortion of the event rate as a 
function of time. This has 
been studied extensively before in great detail and the limit that 
could be put on the $\nutau$ mass ranged from $10~eV$ to $200~eV$ 
\cite{bv}. 

The postbounce supernova neutrinos are emitted in all the three flavors with 
$\numu/\nutau$ $ (\anumu/\anutau)$ having average energies greater than 
$\nue (\anue)$. Non-zero neutrino mass and mixing results in more energetic 
$\numu/\nutau (\anumu/\anutau)$ getting converted to $\nue (\anue)$ thereby 
hardening their resultant energy spectrum and hence enhancing their 
signal at the detector \cite{bkg2,qf,cmk}. 
In a previous work \cite{cmk} we studied quantitatively the effects 
of neutrino 
flavor oscillations on the supernova neutrino spectrum and the number of 
events at the detector. In this work we make a comparative study 
of the neutrino signal 
in the water Cerenkov detectors for a mass range of the neutrinos 
when both the phenomenon of delay and flavor conversion are operative.

We point out that since the time delay of the massive neutrinos is energy 
dependent, and since neutrino flavor conversions change the energy spectra 
of the neutrinos, the time dependence of the event rate at the detector 
is altered appreciably in presence of mixing. For the mass and mixing 
parameters we consider two scenarios and substantiate our point by 
presenting the ratio of the charged to neutral current event rate as a 
function of time for the different cases. We also study the behavior of 
the prompt burst neutrinos in presence of delay and mixing. 
Our study also includes 
the cases of neutrinos with inverted mass hierarchy and degenerate 
neutrinos and we comment on the effect on the observed signal for 
these cases. In section II we briefly describe our model and make predictions 
for the neutrino signal at the detector for massless neutrinos. In section III 
we introduce two different mass and mixing scenarios and make a  comparative 
analysis of the cases (a) with delay effects but with zero mixing and (b) 
with delay along with neutrino flavor conversion. In section IV we consider 
the effect on the event rate for (a) inverted mass hierarchies and (b) 
almost degenerate neutrinos. In section V we discuss our results and 
finally present out conclusions. 


\section {The signal at the detector}

The differential number of neutrino events at the detector for a 
given reaction process is 
\begin{equation}
\frac{d^2 S}{dEdt}=\sum_i\frac{n}{4\pi D^2} N_{\nu_i}(t) \sigma (E)f_{\nu_i}(E) 
\label{sig}
\end{equation}
where $i$ runs over the neutrino species concerned in the given process. 
One uses for the number of neutrinos produced at the source 
$N_{\nu_i}(t) = L_{\nu_i}(t)/\langle E_{\nu_i}(t)\rangle$ 
where $L_{\nu_i}(t)$ is the neutrino luminosity and 
$\langle E_{\nu_i}(t) \rangle$ is the average energy. In eq.(\ref{sig}) 
$\sigma (E)$ is the reaction cross-section for the neutrino with 
the target particle, $D$ is the distance of the neutrino source 
from the detector (taken as 10kpc), 
$n$ is the number of detector particles for the reaction considered  
and $f_{\nu_i} (E)$ is the energy spectrum for the neutrino species involved. 
By integrating out the energy from eq.(\ref{sig}) we get the time 
dependence of the various reactions at the detector. To get the total 
numbers both integrations over energy and time has to be done. We see that 
the eq.(\ref{sig}) has dependence on two things, 
the supernova and the detector. 

{\bf The Supernova:-} 
While for a quantitative analysis people often use numerical supernova model, 
in this work here we adopt a different point of view. It is seen that 
among the supernova models there are uncertainties in the inputs and the 
processes involved and there are differences in the results of  
the different models as well. We have here considered a profile of the 
neutrino luminosities and temperatures which have general agreement, 
though maybe differing in details, with most 
supernova models. The results that we get are not so sensitive to the 
details of model predictions and hence are adequate for our purpose. 

Most of the numerical supernova models agree that the neutrinos carry 
away a few times $10^{53}$ ergs of energy. In this paper we have considered 
that the total energy radiated by the supernova in neutrinos is 
$3\times 10^{53}$ ergs. This luminosity, which is almost the same 
for all the neutrino species, has a fast rise over a period of 0.1 sec 
followed by a slow fall over several seconds. We use a luminosity that 
has a rise of 0.1 sec using one side of the Gaussian with $\sigma$ = 0.03 
and then an exponential decay with time constant $\tau$ = 3 sec. We 
have also incorporated the prompt neutrino burst which occurs when 
the supernova shock crosses the neutrinosphere resulting in the 
emission of a pure electron neutrino pulse created through electron captures. 
Again though there are uncertainties in the numerical models 
regarding the time dependence and the spectrum of this initial $\nue$ 
burst, but in general it has a width of $\sim 10^{-2}$ sec, extending 
between $t\sim 0.04-0.05$ sec postbounce (all times referred to in this 
paper are the time after bounce) and a peak height of about $5\times 
10^{53}$ ergs/sec. The prompt burst neutrinos carry definite energies 
which reflect the energy spectrum of the captured electrons, which in turn 
is a Fermi-Dirac with a certain temperature and chemical  potential. But 
to demonstrate the effect, we assume the simple picture, 
that all the prompt burst neutrinos arrive with 
a constant energy of 10 MeV.  

The average energies associated with the $\nue, \anue ~\rm{and}~ \numu$ (the 
$\numu, \anumu, \nutau ~\rm{and}~ \anutau$ have the same energy spectra) 
are 11 MeV, 16 MeV and 25 MeV respectively in most numerical models. 
There is again disagreement as 
to whether the neutrino average energies rise or fall with time. We take 
them to be constant in time. We have also checked our calculations with 
linearly time dependent average energies and estimated its effect. 
The neutrino spectrum is taken to be a pure Fermi-Dirac 
distribution  characterized by the 
neutrino temperature alone. While most models predict a deviation 
of the spectrum from a pure black body, the difference is small and 
does not change our conclusions much. 

{\bf The Detectors:-} We apply our method to two important water 
Cerenkov detectors, 
the SK and the SNO. While the former has 50 kton of pure $H_2O$ (
fiducial volume is 32 kton), the latter contains 1 kton of pure $D_2O$ 
surrounded by 8 kton of pure $H_2O$ (fiducial volume of light water 
is 1.4 kton). For 
our calculations here we assume that both the detectors have a threshold 
of 5 MeV above which they detect with 100\% efficiency. In Table 1 we 
give the expected number of events for the various processes involved. All 
the numbers are for a 1 kton of detector mass and to get the actual numbers 
one has to multiply them by the fiducial mass \cite{cross}.  
 
\section {The signal with non-zero mass}

If the neutrinos are massless then the time response at the detector 
reflect just the time dependence of their luminosity function at the 
source, which 
is the same for all flavors and hence the same for the charged current 
and neutral current reactions. If neutrinos have mass $\sim eV$ then 
they move at a speed less than the speed of light and hence 
get perceptibly delayed. 
For a neutrino of mass m (in eV) and energy E (in MeV), the delay 
(in sec) in traveling a distance D (in 10 kpc) is 
\begin{equation}
\Delta t(E) = 0.515{(\frac{m}{E})}^2 D
\label{deltime}
\end{equation}
where we have neglected terms second order in (m/E). 
The time response curve then has contributions 
from both the luminosity and the mass and hence the shape of the curve 
changes. If in addition the neutrinos have mixing as well then we 
have seen \cite{bkg2,qf,cmk} that the charge current signals go up. 
This also has an impact on the event rate as a function of time. 
 
The solar neutrino problem, the atmospheric neutrino anomaly and the 
LSND experiment demand mass and mixing parameters in entirely different 
ranges. The solar neutrino deficit demands \cite{bah} \dm $\sim 6.5 
\times 10^{-11} 
eV^2, \sin^2 2\theta \sim 0.75$ (vacuum oscillation solution) or 
\dm $\sim 5 \times 10^{-6} eV^2, \sin^2 2\theta \sim 5.5 \times 10^{-3}$ 
(non-adiabatic MSW solution), the atmospheric $\numu$ depletion is 
explained by \cite{skatm1,cg} \dm $\sim 0.005 eV^2, \sin^2 2\theta 
\sim 1.0$, while LSND gives \dm $\sim  eV^2, \sin^2 2\theta \sim 10^{-3}$ 
which is compatible with the mass required for the hot dark matter in the 
universe which demands $\sum_i m_{\nu_i}$=few $eV$ \cite{hdm}. 
In addition there are 
constraints coming from r-process 
nucleosynthesis in the ``hot bubble" of the supernova which disfavors 
the region of the parameter space with \dm $> 2eV^2$ and \st$>10^{-5}$ 
\cite{qf1}. 
In order to be able to satisfy all these requirements of the 
mass and mixing parameters one would need at least 4-generations of neutrinos 
and perhaps an inverted mass hierarchy as suggested in \cite{fpq}. 
We here work with just the three active neutrino flavors with $m_\nue < 
m_\numu < m_\nutau$ and for the 
mass and mixing parameters we consider two scenarios.

\subsection{Scenario 1}

Here we set $\Delta m_{12}^2 \sim 10^{-6} eV^2$ consistent 
with the solar neutrino problem and $\Delta m_{13}^2 \approx \Delta 
m_{23}^2 \sim 1-10^4 eV^2$ which is the cosmologically interesting 
mass range. Here the mixing angle $\theta_{12}$ can be constrained from 
the solar neutrino data and hence we take $\sin^2 2\theta_{12} \sim 10^{-3}$. 
For $\theta_{13}$ there is no experimental data to fall back upon, but 
from r-process considerations in the ``hot bubble" of the supernova we can 
restrict $\sin^2 2\theta_{13}\sim 10^{-6}$. The atmospheric neutrino anomaly 
in this case would have to be solved by $\numu-\nu_s$ oscillations where 
$\nu_s$ is a sterile species and the LSND data would need some explanation 
other than neutrino oscillations.
 
In this scenario there will be first a matter enhanced 
$\nue-\nutau$ resonance in the mantle of the supernova followed by a 
$\nue-\numu$ resonance in the envelope. The two resonances are very well 
separated and the mixing angle very small, so that the 
survival probability for an  
emitted $\nue$ of energy $E$ in the Landau-Zener approximation 
is given by \cite{qf,kuo}
\begin{equation}
P_{\nue\nue} \approx \exp\{-\pi(H_{12}\Delta m_{12}^2 \sin^2 2\theta_{12} +
H_{13}\Delta m_{13}^2 \sin^2 2\theta_{13})/4E\}
\label{p1}
\end{equation}
where the $H_{12},~H_{13}$ are the density scale heights at the position 
of the $\nue-\numu$ and $\nue-\nutau$ resonances respectively, given 
by $H=|\rm{dln}\rho/\rm{dr}|_{\rm{res}}^{-1}$. From the density 
profiles of the mantle and 
the envelope of a given supernova model, the survival probability for $\nue$ 
can be calculated from eq.(\ref{p1}). It has been shown in \cite{smir} 
that 
for the \dm considered here, both the resonances are completely adiabatic 
so that there is no level jumping. Even if the jump probability  given 
by eq.(\ref{p1}) is not exactly zero, especially for the $\nue-\numu$ 
resonance, we expect that the effect will be small and hence 
for the extremely small 
mixing angles that we are considering here, we take the survival probability 
$P_{\nue\nue} \approx 0$. 

As a result of this MSW conversion in the supernova, though the $\nue$ 
flux goes down in numbers, it gains in average energy resulting in an 
enhanced signal at the detector \cite{cmk}. Of course here since the $\anue$ 
do not have any conversion, the $\anue$ signal remains unaltered. The 
neutral current events being flavor blind do not show any change either. 
The shape of the $\nue$ charged current event rate as a function of time 
remains the same even though the event rate increases. But the prompt burst 
neutrinos are absent in the charged current signal 
if the $\nue$ get completely converted to $\numu$. 
The total number of events integrated over time 
in this scenario with flavor conversion are given in Table 1.

For massive neutrinos with no mixing we will have only delay effects. 
For the masses assumed here only the $\nutau$ will be delayed. 
The expression for the neutral current event rate in the 
detector is then given by
\begin{eqnarray}
\frac{dS_{nc}^d}{dt}&=&\frac{n}{4\pi D^2} \int dE \sigma (E) 
\{N_\nue(t) f_\nue (E) + N_\anue (t)f_\anue (E) + 
2N_\numu(t) f_\numu (E)
\nonumber\\ 
&+& 2 N_\nutau(t-\Delta t(E))f_\nutau(E)\}
\label{del1}
\end{eqnarray} 
where $dS_{nc}^d/dt$ denotes the neutral current (nc) event rate with 
delay (d). 
Delay therefore distorts the rate vs. time curve. By doing a $\chi^2$ 
analysis of this shape distortion one can put limits on the mass 
that can be detected by either of the detectors \cite{bv}.


If the neutrinos have mass as well as mixing and if we consider 
complete flavor conversion inside the supernova, then since the $\numu$ and 
the $\nutau$ spectra are identical, the eq.(\ref{del1}) reduces to,
\begin{eqnarray}
\frac{dS_{nc}^{do}}{dt}&=&\frac{n}{4\pi D^2} \int dE \sigma (E)
\{N_\nue(t-\Delta t(E))f_\nue(E)+ N_\anue (t)f_\anue (E)
\nonumber\\
&+&2N_\numu(t) f_\numu (E)+N_\nutau(t) f_\nutau (E)+
N_\anutau(t-\Delta t(E))f_\anutau(E)\} 
\label{do1}
\end{eqnarray}
where $dS_{nc}^{do}/dt$ is the reaction rate with delay and flavor 
conversion. 

In fig. 1 we have plotted the neutral current event rate for the reaction 
($\nu_x + d \rightarrow n+p+\nu_x$) as a function of time for massless 
neutrinos along with the cases for mass but no mixing (eq.(\ref{del1})) 
and mass 
along with mixing (eq.(\ref{do1})). We have used a log scale scale for the 
time axis to show the prompt burst neutrinos. 
The figure looks similar for the other 
neutral current reactions as well, apart from a constant normalization factor 
depending on the total number of events of the process concerned. The 
curves corresponding to the massive neutrinos have been given for 
$m_\nutau = 40 eV$. As the delay given by eq.(\ref{deltime}) depends 
quadratically 
on the neutrino mass, the distortion is more for larger masses. From the 
fig. 1 we see that the prompt neutrino burst shows up at a much later 
time in presence of delay and complete mixing. For partial conversion 
($P_{\nue\nue} \neq 0$), the prompt burst neutrinos would break up into 
two, an unconverted $\nue$ component and a massive component, 
depending on the value of $P_{\nue\nue}$ and while the former 
would arrive undelayed at t$\approx0.04 sec$, the latter 
component would get delayed by about 8 sec for $m_\nutau$=40 eV. 
The comparison of the cases with 
and without mixing shows that the distortion of the curve due to delay 
is diluted in the presence of mixing. Although it may seem that the curve 
with delay and mixing can be simulated by another curve with delay alone 
but with smaller mass, the actual shape of the two curves would still be 
different. This difference in shape though may not be statistically 
significant for the present water Cerenkov detectors. 

In fig. 2 we give the ratio R(t) of the total charged current to the 
neutral current event rate in SNO as a function of time. Plotted are the 
ratios (i) without mass, (ii) with only mixing, (iii) with delay 
but zero mixing and (iv) with delay 
and flavor mixing. The differences in the behavior of R(t) for the four 
different cases are clearly visible. For no mass R(t)=0.3 and since the 
time dependence of both the charged current and neutral current reaction 
rates are the same, their ratio is constant in time. With only mixing 
the ratio increases appreciably and goes upto R(t)=0.61, remaining constant 
in time, again due to the same reason. With the introduction of delay the 
ratio becomes a function of time as the neutral current reaction now has 
an extra time dependence coming from the mass. At early times as the 
$\nutau$ get delayed the neutral current event rate drops increasing 
R(t). These delayed $\nutau$s arrive later and hence R(t) falls at 
large times. 
This feature can be seen for both the curves with and without mixing. 
The curve for only delay starts at R(t)=0.52 at t=0sec and falls to about 
R(t)=0.26 at t=10sec. For the delay with mixing case the corresponding 
values of R(t) are 0.83 and 0.51 at t=0 and 10 sec respectively. 
The noteworthy thing is that the curves with and without mixing are 
clearly distinguishable and should allow one to differentiate between the 
two cases of only delay and delay with neutrino flavor conversion.

\subsection{Scenario 2}

Here we set $\Delta m_{12}^2 \approx \Delta m_{13}^2 
\sim 1-10^4 eV^2$ corresponding to the cosmological range and 
$\Delta m_{23}^2 \sim 10^{-3} eV^2$ which will explain the atmospheric 
neutrino anomaly. As before we take the mixing angle to be small from 
r-process considerations. Here we assume that it is the $\nue-\nu_s$ 
mixing which is responsible for the solar neutrino deficit. Though this 
scheme can accommodate LSND in principle, we will generally consider masses 
that are higher than the range over which LSND is sensitive. 

For a mass spectrum like this, there will be just 
one matter enhanced resonance in the supernova instead of two as has 
been numerically shown in \cite{zs}. Hence the survival probability 
for $\nue$ of energy $E$ in the Landau-Zener approximation is given by 
\begin{equation}
P_{\nue\nue} \approx \exp\{-\pi(H_{13}\Delta m_{13}^2 \sin^2 
2\theta_{13})/4E\}\
\end{equation}
and as argued before for adiabatic conversion with small mixing angles
we take $P_{\nue\nue} \approx 0$. The charged current event rate as a 
function of time and the 
time integrated number of events in 
this scenario therefore remains the same as in scenario 1 (see column 
3 of Table 1). 

For this case both the $\numu$ and the $\nutau$ will be delayed and so the 
delay effects will be almost twice that we had for the scenario 1. The 
neutral current time response in the detector for this mass scenario 
without mixing  is given by
\begin{eqnarray}
\frac{dS_{nc}^d}{dt}&=&\frac{n}{4\pi D^2} \int dE \sigma (E) 
\{N_\nue(t) f_\nue (E) + N_\anue (t)f_\anue (E)  
\nonumber\\ 
&+& 4 N_\nutau(t-\Delta t(E))f_\nutau(E)\}
\label{del2}
\end{eqnarray} 
From the expressions (\ref{del1}) and (\ref{del2}) it is clear that the 
event rate in scenario 2 with a certain mass $m$ for both $\numu$ and 
$\nutau$ is not exactly equal to the reaction rate in scenario 1 with a 
$\nutau$ mass of $2m$. 

If the neutrinos have mass as well as mixing then eq.(\ref{del2}) 
for complete conversion $P_{\nue\nue}\approx 0$ becomes 
\begin{eqnarray}
\frac{dS_{nc}^{do}}{dt}&=&\frac{n}{4\pi D^2} \int dE \sigma (E)
\{N_\nue(t-\Delta t(E))f_\nue(E)+ N_\anue (t) f_\anue (E)
\nonumber\\
&+&2N_\numu(t-\Delta t(E)) f_\numu (E)+N_\nutau(t)f_\nutau(E)
+N_\anutau(t-\Delta t(E)) f_\anutau (E)\} 
\label{do2}
\end{eqnarray}

In fig. 3 we plot as in scenario 1, the ($\nu_x-d$) event rate as a 
function of time, for $m_\numu=m_\nutau=40eV$. Clearly the 
delay effects are more in this case than in scenario 1. 
The prompt burst neutrinos have a behavior similar to that seen in the 
scenario 1. 
In fig. 4 we give the ratio R(t) of the charged current 
to neutral current reaction rates in SNO as before. All the features 
that were present earlier are also present here. For no mass R(t)=0.3, 
which rises to 0.61 on introduction of complete flavor conversion, 
being constant in time 
as before. But here the distortion of the shape of R(t) due to delay 
is immense. With the introduction of delay the ratio rises to  
about 1.85 at t=0sec and then falls sharply with time to  
0.27 by t=2sec, while for the case of delay with flavor mixing the ratio 
falls from 1.95 at t=0sec to about 0.57 at t=2sec. Again here the two curves 
are clearly distinguishable. Also the shape of the R(t) curves in the 
first 2 seconds are different from the shape of the R(t) curves in the 
first 2 seconds of scenario 1. This should help us to the distinguish 
between the two mass patterns considered in scenario 1 and 2. 

\section{Signal with other mass patterns}

Apart from the ones that we have considered in the previous section, there 
may be other mass hierarchies for the neutrinos as well. We can have a 
situation where the neutrino state corresponding to the $\nue$ is the 
heaviest mass state \cite{fpq,inv}. Such ``inverted" mass hierarchies 
do not interfere with the r-process in the 
``hot bubble" \cite{qf2} though they maybe in conflict with the SN1987A 
data. The SN1987A analysis of Smirnov, Spergel and Bahcall in \cite{smir} 
constrains $P_{\nue\nue} \geq 0.65$. 
But it has been argued by Fuller, Primack and Qian 
\cite{fpq} that the constraints from SN1987A are not strong enough and 
so one may have large resonant antineutrino conversions in the 
supernova. In Table 1 we report the total signal at the detector for the 
case $m_{\nu_1} \ll m_{\nu_2} \sim m_{\nu_3}$  
with $P_{\nue\nue}$ = 0 in column 4 and with $P_{\nue\nue}$ = 0.65 in 
column 5.

If $\nue$s were massive they would get delayed and we would see delay 
effects in the charged current signal as well. This would create an 
interesting situation in which the conventional methods suggested in the 
literature to determine the neutrino mass from the time of flight 
measurements would fail. But the $^3H 
\beta$ decay experiments restrict $m_\nue < 5 eV$ \cite{balasev} (in fact 
the bound may be even lower \cite{lobasev}). Since the time delay given 
by eq.(\ref{deltime}) depends quadratically on the neutrino mass, the 
delay effects for such low $m_\nue$ mass is very small. The only  
signature in this case maybe in the ratio R(t) of the charged current to 
neutral current event rate in the first 0.2 sec. In the fig. 5 we plot the 
ratio R(t) in SNO for the various cases as a function of time for  
$m_\numu \ll m_\nutau \sim m_\nue = 5 eV$ and with $P_{\nue\nue}$=0. 
For the no mass case R(t)=0.3 which rises to 0.5 when mixing is switched 
on. For the only delay case the ratio R(t) rises from almost 0.02 initially to 
about 0.3 at t=0.2sec, the no mass value, beyond which it remains 
constant. For the case of delay with mixing R(t) rises from 0.25 at t=0.01sec 
to about 0.5 at 
t=0.2sec, the only mixing value and remains constant thereafter. 
The prompt burst neutrinos get delayed in this case 
even though they do not have flavor conversion here and appear at the same 
delayed time for charged current and neutral current events. 

Instead of hierarchical spectrum, neutrinos may even be almost degenerate 
with a few eV mass \cite{deg}. The small mass splitting between them 
can account for the solar neutrino problem (\dm $\sim 10^{-6} eV^2$) 
and the atmospheric neutrino anomaly (\dm $\sim 10^{-3} eV^2$). This mass 
pattern does not contradict the r-process as the mass 
differences here are small and not enough to cause neutrino conversions in 
the ``hot bubble". This scheme of course cannot account for the LSND 
results. Again here we may have either $m_\nue < m_\numu < m_\nutau$ or 
$m_\nue > m_\numu > m_\nutau$. While for the former we will have 
$\nue-\nutau$ resonance followed by $\nue-\numu$ resonance, for the latter 
the antineutrinos will similarly resonate. From the expression (\ref{p1}) 
one can calculate the relevant probabilities. In this case, for both normal 
as well as inverted masses, all the three neutrino species will be delayed. 
Hence here also we can expect delay effects in both the charged as well as 
the neutral current events. But here again since $m_\nue$ cannot be larger 
than 5 eV, the other species will also have a maximum 
mass of about 5 eV and the 
delay effects are not large. But again one can expect to see some change 
in the ratio R(t) of the charged to neutral current event rates as a function 
of time. In figs. 6 and 7 we plot the ratio R(t) for 
 $m_\nue < m_\numu < m_\nutau$ and $m_\nue > m_\numu > m_\nutau$ 
respectively with $P_{\nue\nue}\approx 0$ for the cases with mixing. 
The prompt neutrino burst appears delayed in both the charged 
current as well as neutral current signal for the cases where 
$m_\nue > m_\numu > m_\nutau$. For the $m_\nue < m_\numu < m_\nutau$ 
the $\nue$ are transformed to $\numu$ and do not appear in the charged 
current signal, though they do appear at a delayed time in the neutral 
current events.

\section{Discussions and Conclusions}

As the average energies of the $\numu/\nutau(\anumu/\anutau)$ are greater 
than the average energies of the $\nue(\anue)$, neutrino flavor mixing 
modifies the energy spectrum of the neutrinos. As the detection 
cross-sections are highly energy dependent this results in the 
enhancement of the charged current signal, but as the neutral current 
reactions are flavor blind, the total neutral current signal remains 
unchanged. However the time delay $\propto 1/E^2$, and as the energy 
spectrum of the neutrinos change the resultant delay is also modified and 
this in turn alters the neutral current event rate as a function of time. 
MSW conversion in the supernova results in de-energising the $\numu/\nutau$ 
spectra and hence the delay effect should increase. As larger delay caused 
by larger mass results in further lowering of the neutral current event 
rate vs. time curve for early times, one would normally expect that the 
enhanced delay as a result of neutrino flavor conversion would have a 
similar effect. But the figs. 1 and 3 show that the curves 
corresponding to delay with mixing are higher than the ones with only time 
delay. This at first sight seems unexpected. But the point to note is 
that while the flavor conversion reduces the average energies of the massive 
$\nu_x$ (where $x$ stands for $\mu/\tau$) increasing its delay and hence 
depleting its signal at early times, it energizes the $\nue$ beam which 
is detected with full strength. Therefore while for no mixing the $\nu_x$ 
gave the largest signal, for the case with mixing it is the $\nue$ that 
assume the more dominant role. If we consider the scenario 1, for the only 
delay case the delayed $\nutau$s appear after t$\approx 0.1 sec$, 
while for delay with flavor conversion the $\nutau$s appear only after 
t$\approx 0.5 sec$. So upto t$\approx 0.1 sec$, for only delay we have 
just the $\nue$ signal with $\langle E_\nue \rangle =11 MeV$ 
while for delay with 
mixing we have $\nue$ signal with $\langle E_\nue \rangle =25 MeV$. 
Hence for t$<0.1 sec$ the signal for delay with mixing is more 
as the signal goes up significantly for higher energy neutrinos. Beyond 
t$\approx 0.1 sec$ the delayed $\nutau$s start appearing for the only delay 
case and by t$\approx 0.7 sec$ the signal for only delay becomes equal 
to the signal for delay with mixing. By this time the delayed $\nutau$s 
with $\langle E_\nutau\rangle =11 MeV$ also appear for the 
delay with mixing case 
but their contribution is small. By $t\approx 1 sec$ the two curves 
corresponding to to only delay and delay with mixing approach each 
other and the difference between them becomes hard to detect.  
 
For scenario 1 and 2 the $\anue$ 
spectrum remains unchanged. Hence we can use the large $\anue$-p signal 
at SK to fix the supernova model parameters as a function of time. This 
can then be used to make predictions for the $\nue$ charged current signal. 
A comparison of the expected to the actual $\nue$ signal would give the 
value of $P_{\nue\nue}$. This $P_{\nue\nue}$ can then be incorporated in 
the analysis and a correct limit for the neutrino mass from the neutral 
current event rate can be obtained. 

On the other hand if the $\anue$ captures are much larger than most model 
predictions, then that would indicate a $\anue$ MSW resonance in the 
supernova and hence an inverted mass hierarchy. In that case, 
in principle  we would see 
delay effects in both the neutral as well as the charged current events. 
But the current laboratory limits on $\nue$ mass ($m_\nue \leq 5 eV$) ensure 
that these effects may not be discernible in the present detectors. The 
neutral current to charged current ratio in the first 0.2 sec could still 
carry some clue regarding neutrino mass. The other quantity that could in 
principle be searched for is the prompt neutrino burst. This will appear, 
if detected, at the same delayed time in the charged and the neutral 
current signal.   
  
One parameter which carries both the information of the neutrino mass and 
their mixing is R(t), the ratio of charged to neutral current event rate as 
a function of time. This ratio as a tool to look for mixing has been 
suggested before. We have shown here that the shape of R(t) changes in 
presence of time of flight delay. This shape distortion can be used to put 
limits on neutrino mass. The ratio is not only sensitive to mass and 
mixing parameters it is also almost model independent. It is almost 
independent of the luminosity and depends only 
on some function of the ratio of neutrino temperatures. 
We have also repeated our analysis 
with a linear time dependence for the neutrino temperature and have 
found that the time dependence of the neutrino temperature does not have 
much effect on the time dependence of the event rates or the ratio of the 
charged current to neutral current rates. 

We have also studied the prompt neutrino burst under the different possible 
mass and mixing schemes where one can have perceptible time delay. For 
normal mass hierarchies with mass and mixing the prompt burst neutrinos 
are delayed in the neutral current signal and are absent in the $\nue$ charged 
current signal. For inverted mass hierarchies the $\nue$s are delayed but 
do not undergo flavor conversion and so the prompt burst neutrinos come at the 
same delayed time for both charged as well as neutral current signal. 

In conclusion, we point out that one should consider the delay effects 
of massive neutrinos with mixing as it has been proved beyond doubt that 
the neutrinos do mix. 
We have shown here that neutrino mixing leads to 
nontrivial changes in the time response of the neutrinos at the detector. 
The prompt burst neutrinos and the charged current to neutral current 
ratio are important tools which carry information about the neutrino mass 
and mixing as demonstrated here and should be studied carefully.

\vspace{1cm}

{\small The authors wish to thank J.Beacom and S.Goswami for helpful 
discussions. KK acknowledges helpful suggestions from G. Rajsekaran, 
M.V.N. Murthy and D. Indumathi.

\newpage

\begin{description}
\item{{\bf Table 1}} The expected number of neutrino events for a 1 kton  
water Cerenkov detector. The column A corresponds to massless neutrinos, 
column B to neutrinos with normal hierarchies ($m_\nue < m_\numu < m_\nutau$) 
and complete flavor conversion 
and column C1 and C2 to neutrinos with inverted mass hierarchies 
($m_\nue > m_\numu > m_\nutau$). C1 is for $P_{\nue\nue}=0$ while 
C2 is for $P_{\nue\nue}=0.65$, the lower allowed limit from SN1987A 
\cite{smir}. The $x$ here refers to all the six neutrino species. 
\end{description}
\[
\begin{tabular}{|c|c|c|c|c|} \hline 
{reaction} & {A} & {B} & {C1} & {C2}\\ \cline{1-5}
\hline


{$\nu_e+d\rightarrow p+p+e^-$} & { 75 } & {239} 
& {75} & {75}\\ \hline
{$\bar\nu_e +d\rightarrow n+n+e^+$} & {91} & {91} 
& {200} & {129}\\ \hline
{$\nu_x+d\rightarrow n+p+\nu_x$} & {544} & {544} & {544} & 
{544} \\ \hline
{$\bar\nu_e +p\rightarrow n+e^+$} & {255} & {255} 
& {423} & {314}\\ \hline
{$\nu_e +e^- \rightarrow \nu_e +e^-$} 
& {4.45} & {6.54} & {4.45} & {4.45} \\ \hline
{$\bar\nu_e+e^- \rightarrow \bar\nu_e+e^-$} 
& {1.54} & {1.54} & {2.09} & {1.73} \\ \hline
{$\nu_{\mu,\tau}(\bar\nu_{\mu,\tau}) +e^- \rightarrow 
\nu_{\mu,\tau}(\bar\nu_{\mu,\tau}) +e^-$} 
& {4.13} & {3.78} & {3.96} & {4.07} \\ \hline
{$\nu_e +^{16}O \rightarrow e^- +^{16}F$} & {1.39} & {61.60} 
& {1.39} & {1.39} \\ \hline
{$\bar\nu_e + ^{16}O\rightarrow e^+ + ^{16}N$} & {4.23} & {4.23} 
& {20.60} & {9.96} \\ \hline
{$\nu_x +^{16}O \rightarrow \nu_x +\gamma +X$} & {23.4} & {23.4} & 
{23.4} & {23.4} \\ \hline 
\end{tabular}
\]

\newpage

\begin{center}
{\bf Figure Captions}
\end{center}

\noindent
Fig.1 The event rate as a function of time for the reaction ($\nu_x$-d) 
in SNO for the scenario 1. The solid line corresponds to the case of 
massless neutrinos, the long dashed line to neutrinos with only mass but 
no mixing, while the short dashed line gives the event rate for neutrinos 
with mass as well as complete flavor conversion. The prompt burst neutrinos, 
undelayed for the no mixing case and delayed for the case of delay and 
mixing can be seen.

\vskip 0.5 cm

\noindent
Fig.2 The ratio R(t) of the total charged current to neutral current 
event rate 
in SNO versus time for the scenario 1. The solid line is for massless 
neutrinos, the long dashed line for neutrinos with only flavor conversion but 
no delay, the short dashed line for neutrinos with only delay and no flavor 
conversion and the dotted line is for neutrinos with both delay and flavor 
conversion. The sharp dips in the ratio is due to conversion of the 
prompt burst $\nue$ to $\numu$ when mixing is taken into account, 
which then are absent in the charged current 
but are present in the neutral current signal. The delayed and converted 
prompt $\nue$ burst is also visible. 

\vskip 0.5 cm

\noindent
Fig.3 Same as in Fig.1 but for the scenario 2.

\vskip 0.5 cm

\noindent
Fig.4 Same as in Fig.2 but for the scenario 2.

\vskip 0.5 cm

\noindent
Fig.5 Same as in Fig.2 but for neutrinos with inverted mass hierarchy 
with $m_{\nu_1}\ll m_{\nu_2} \sim m_{\nu_3} = 5 eV$. 

\vskip 0.5 cm

\noindent
Fig.6 Same as in Fig.2 but for almost degenerate neutrinos ($m_{\nu_i}=5eV$) 
with $m_\nue < m_\numu < m_\nutau$.

\vskip 0.5 cm

\noindent
Fig.7 Same as in Fig.6 but for $m_\nue > m_\numu > m_\nutau$.

\vskip 0.5 cm


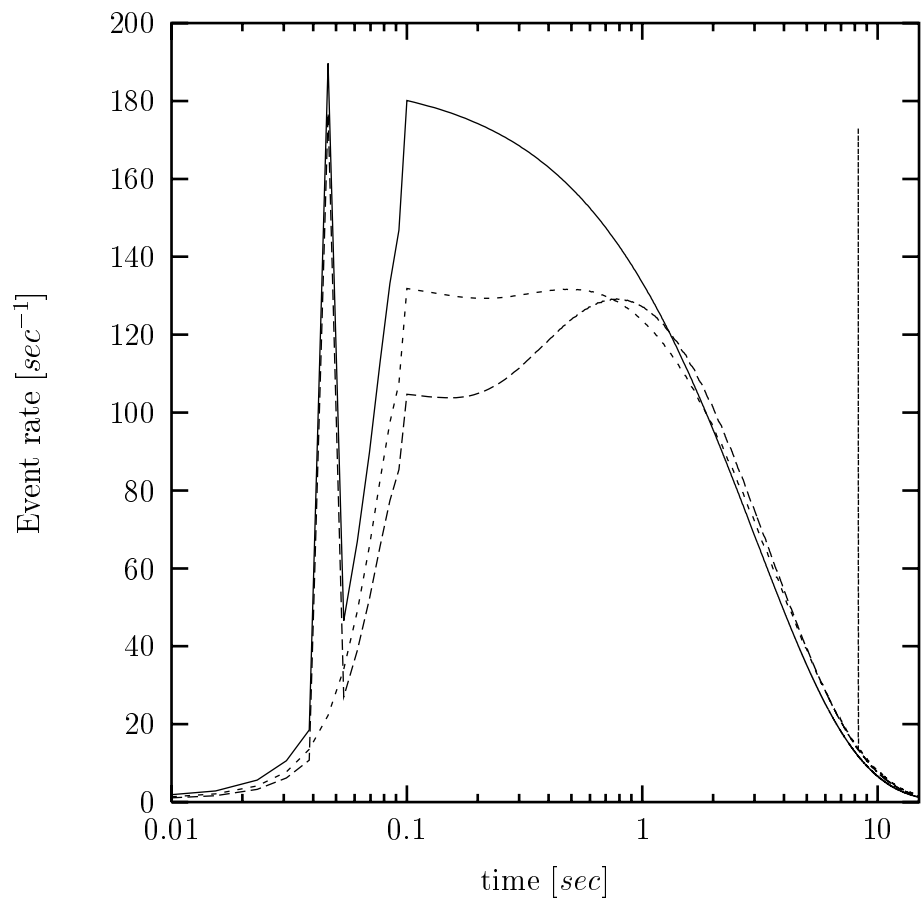
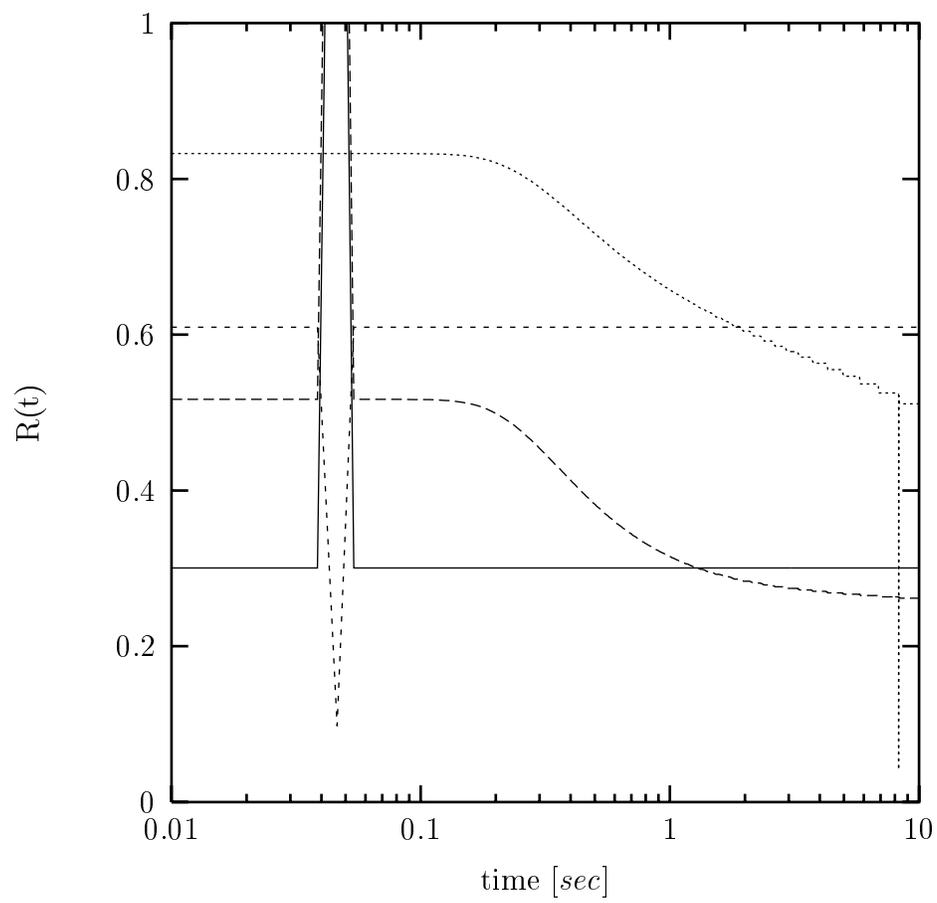
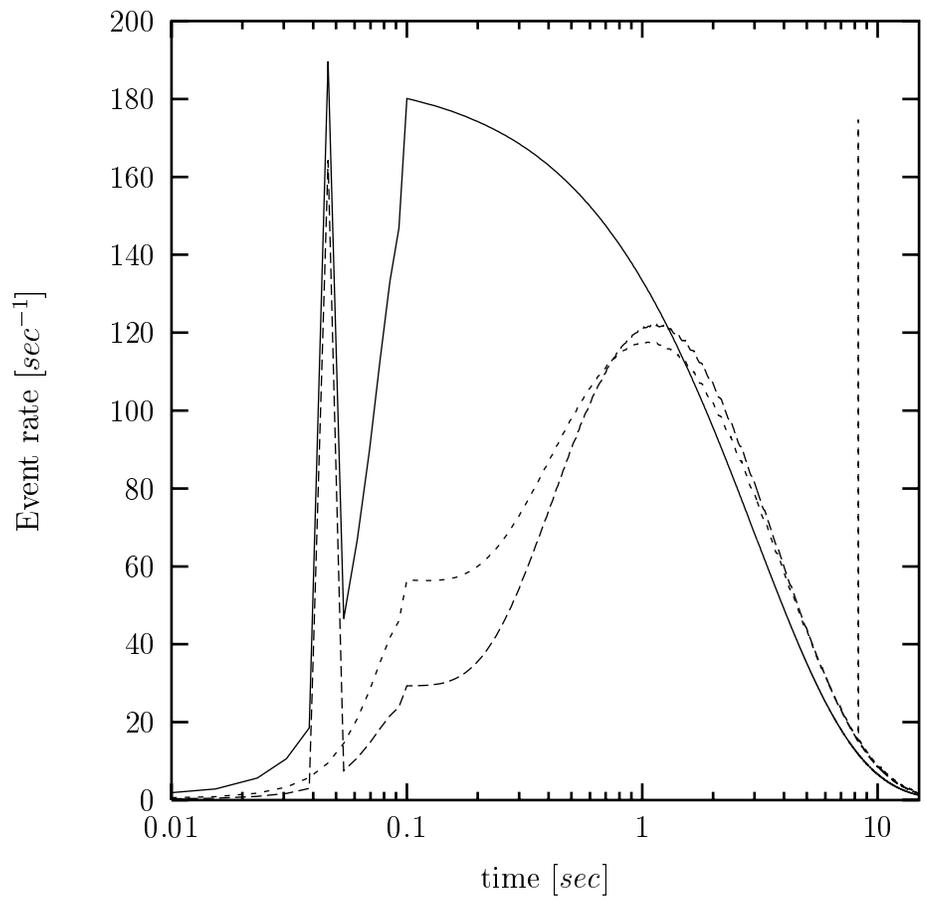
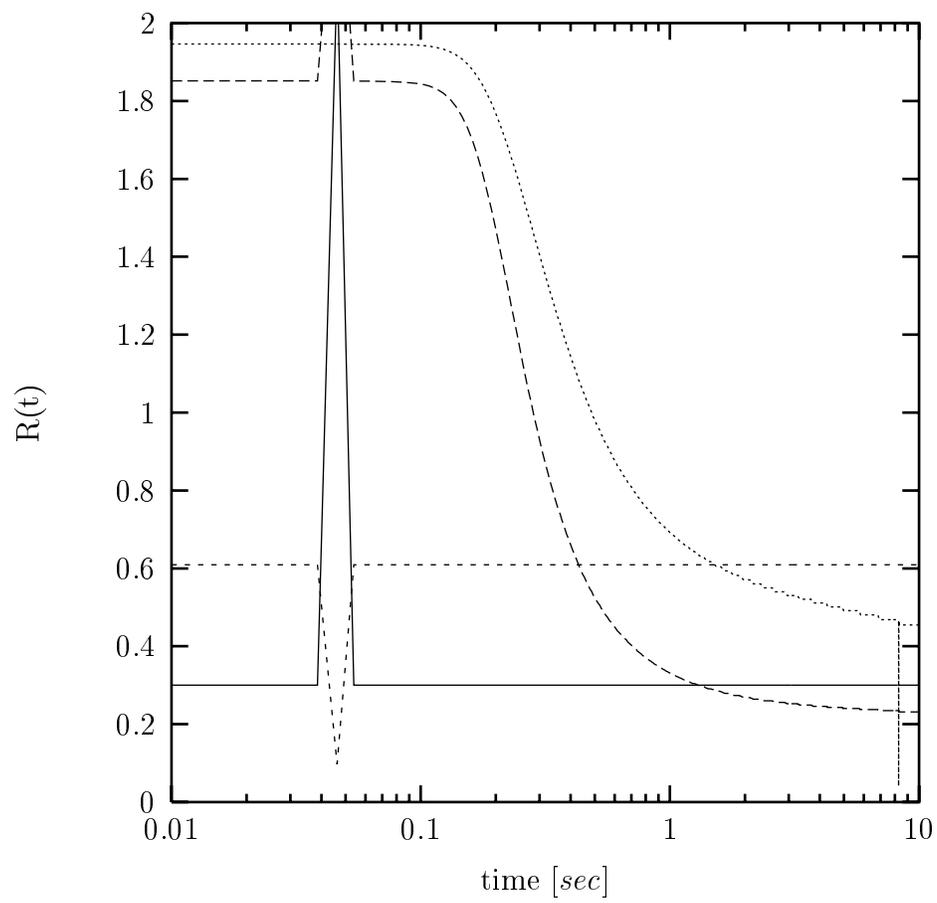
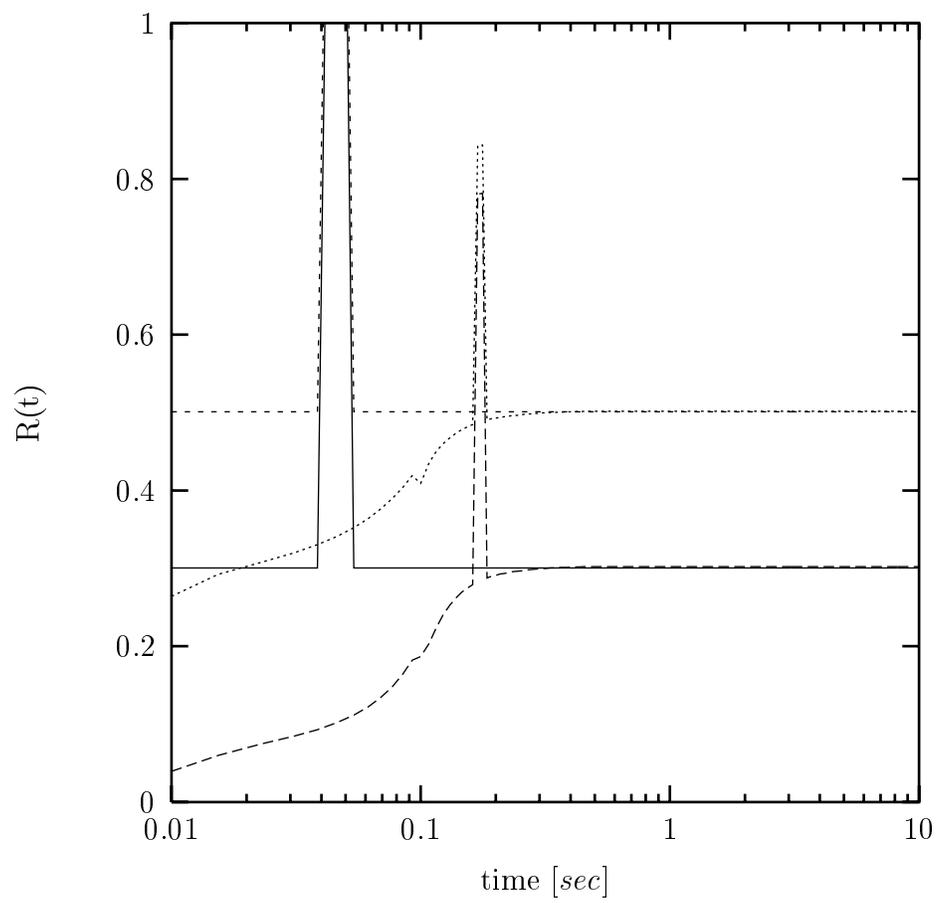
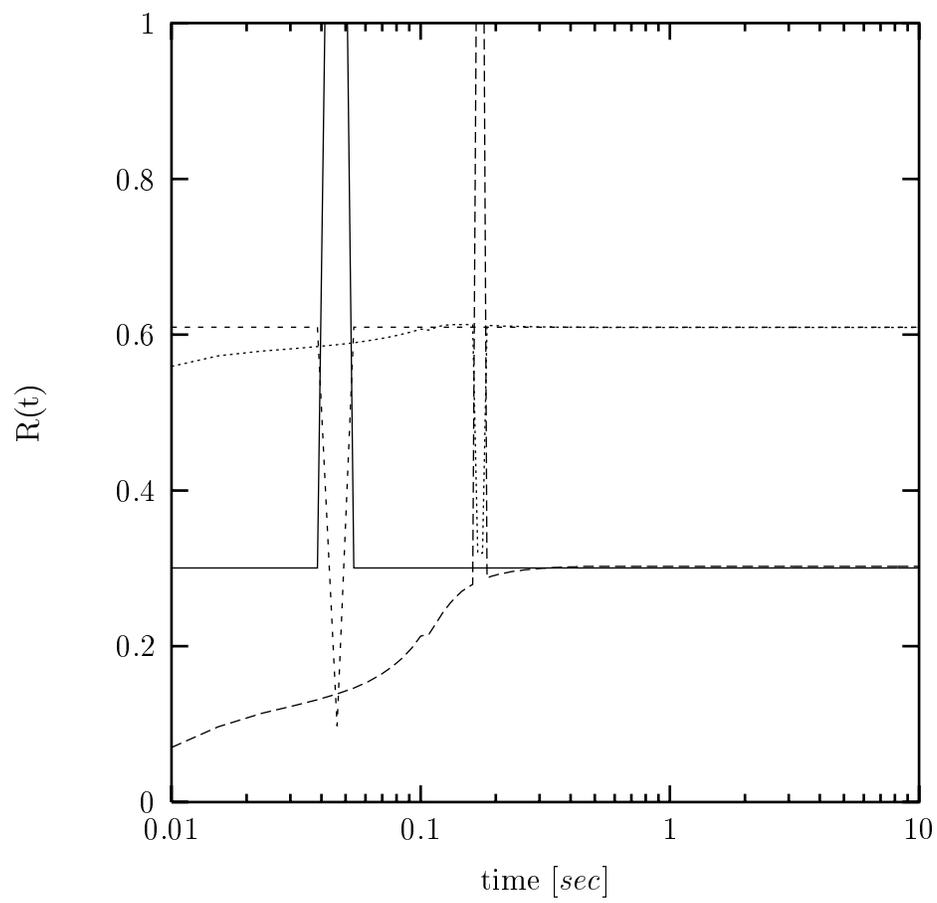
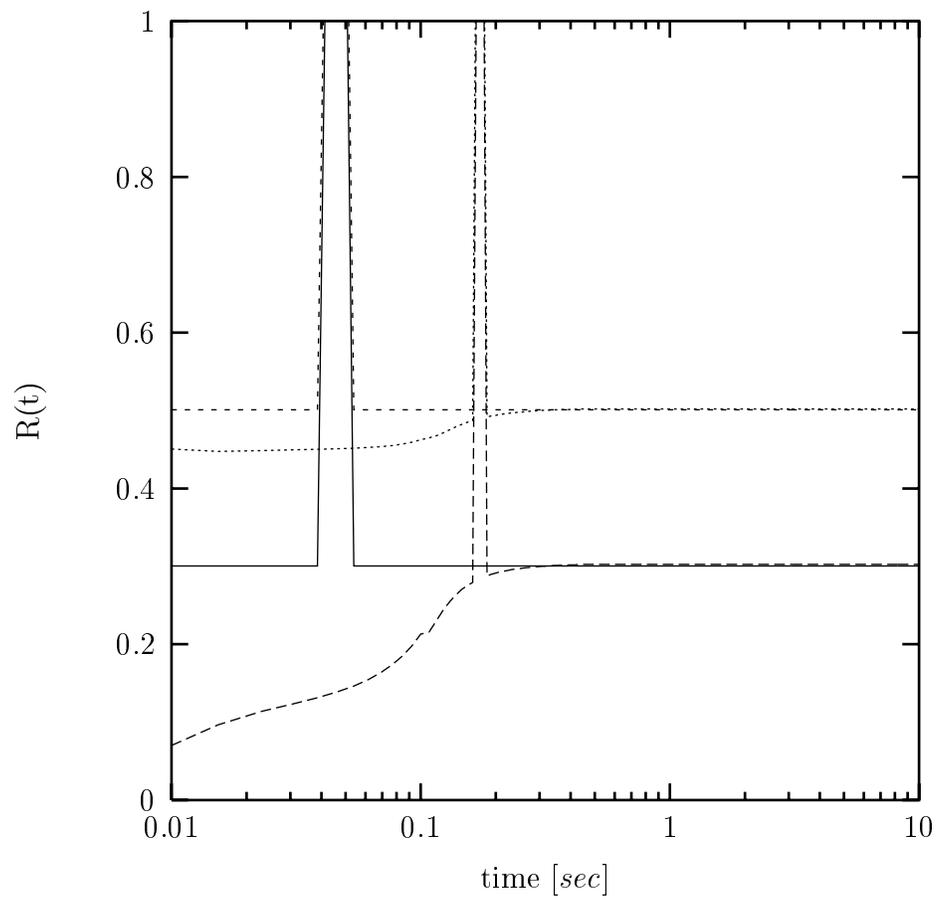
\end{document}